\documentstyle[prd,aps]{revtex}
\begin{document}
\draft
\baselineskip 16pt
\renewcommand{\topfraction}{0.8}
\renewcommand{\textwidth}{15cm}
\renewcommand{\topmargin}{-0.5cm}
\preprint{OU-TAP 142}
\tighten
\title{Inflation and the dwarf galaxy problem %
}

\author{ Jun'ichi YOKOYAMA}
\address{\hfill\\
Department of Earth and Space Science, Graduate School of Science,\\
Osaka University, Toyonaka 560-0043, Japan\\
}

\maketitle

\abstract{Cosmological $N$-body simulations with scale-invariant initial
perturbations show too many virialized objects of dwarf-galaxy masses in
a typical galactic halo in contradiction to observation and stability of
galactic discs.  We present a model of double hybrid inflation which predicts density
fluctuations with a small-scale cutoff to account for the absence 
of these objects in observations.
}


\maketitle

%

\newcommand{\dw}{{\rm DW}}
\newcommand{\cw}{{\rm CW}}
\newcommand{\ml}{{\rm ML}}
\newcommand{\lt}{\tilde{\lambda}}
\newcommand{\lh}{\hat{\lambda}}
\newcommand{\phidot}{\dot{\phi}}
\newcommand{\phicl}{\phi_{cl}}
\newcommand{\adot}{\dot{a}}
\newcommand{\phat}{\hat{\phi}}
\newcommand{\ahat}{\hat{a}}
\newcommand{\hhat}{\hat{h}}
\newcommand{\phihat}{\hat{\phi}}
\newcommand{\Nhat}{\hat{N}}
\newcommand{\hth}{h_{th}}
\newcommand{\hbh}{h_{bh}}
\newcommand{\gsim}{\gtrsim}
\newcommand{\lsim}{\lesssim}
\newcommand{\bfx}{{\bf x}}
\newcommand{\bfy}{{\bf y}}
\newcommand{\bfr}{{\bf r}}
\newcommand{\bfk}{{\bf k}}
\newcommand{\bkp}{{\bf k'}}
\newcommand{\order}{{\cal O}}
\newcommand{\beq}{\begin{equation}}
\newcommand{\eeq}{\end{equation}}
\newcommand{\beqa}{\begin{eqnarray}}
\newcommand{\eeqa}{\end{eqnarray}}
\newcommand{\mpl}{M_{Pl}}
\newcommand{\lmk}{\left(}
\newcommand{\rmk}{\right)}
\newcommand{\lkk}{\left[}
\newcommand{\rkk}{\right]}
\newcommand{\lnk}{\left\{}
\newcommand{\rnk}{\right\}}
\newcommand{\call}{{\cal L}}
\newcommand{\calr}{{\cal R}}
\newcommand{\half}{\frac{1}{2}}
\newcommand{\kc}{\kappa\chi}
\newcommand{\bkc}{\beta\kappa\chi}
\newcommand{\gkc}{\gamma\kappa\chi}
\newcommand{\gbkc}{(\gamma-\beta)\kappa\chi}
\newcommand{\dchi}{\delta\chi}
\newcommand{\dphi}{\delta\phi}
\newcommand{\dOmega}{\delta\Omega}
\newcommand{\Phibd}{\Phi_{\rm BD}}
\newcommand{\echi}{\epsilon_\chi}
\newcommand{\ephi}{\epsilon_\phi}
\newcommand{\Phihat}{\hat{\Phi}}
\newcommand{\Psihat}{\hat{\Psi}}
\newcommand{\that}{\hat{t}}
\newcommand{\Hhat}{\hat{H}}
\newcommand{\zk}{z_k}
\newcommand{\msolar}{M_\odot}
\newcommand{\mbh}{M_{\rm BH}}
\newcommand{\bh}{{\rm BH}}
\newcommand{\calf}{{\cal F}}
\newcommand{\gtilde}{{\tilde g}}
\newcommand{\lcdm}{$\Lambda$CDM}
\newcommand{\chione}{{\chi_1}}
\newcommand{\chionem}{\chi_{1m}}
\newcommand{\chitwo}{{\chi_2}}
\newcommand{\phione}{{\phi_1}}
\newcommand{\phionec}{\phi_{1c}}
\newcommand{\phitwo}{{\phi_2}}
\newcommand{\phitwom}{\phi_{2m}}
\newcommand{\phitwoc}{\phi_{2c}}
\newcommand{\Vone}{V_1}
\newcommand{\Vtwo}{V_2}
\newcommand{\vone}{v_1}
\newcommand{\vtwo}{v_2}
\newcommand{\mone}{m_1}
\newcommand{\mtwo}{m_2}
\newcommand{\lone}{\lambda_1}
\newcommand{\ltwo}{\lambda_2}
\newcommand{\gone}{g_1}
\newcommand{\gtwo}{g_2}
\newcommand{\alphaone}{\alpha_1}
\newcommand{\alphatwo}{\alpha_2}

\section{Introduction}

Inflationary cosmology \cite{oriinf,inf} 
has strongly influenced the study of cosmological
structure formation for these two decades, because its simplest versions
predict a rather specific universe with a vanishingly small spatial curvature
and scale-invariant adiabatic fluctuations \cite{fluc}.  
The standard models of structure formation today 
starts with these initial conditions and assume that dominant material contents
is cold dark matter (CDM) and a
cosmological-constant-like energy of unknown origin.  

The standard
$\Lambda$CDM model has proved remarkably successful in explaining
various observational data.
In this model the gravitational clustering process is governed by
the dark matter component and the baryons do not play important roles.
Since gravity is a scale-free force by nature, we would end up with a
hierarchical universe if we started with the scale-invariant 
primordial fluctuations.  In such a universe, galaxy halos would 
appear as scaled versions of clusters of galaxies.  That is, just as
clusters are associated with thousands of substructure clumps with a steep 
mass spectrum, we expect many virialized substructures of 
dwarf-galaxy-scales in galactic halos.  
Indeed, such substructures are shown to be
produced on the relevant scale by recent $N$-body simulations with higher
resolution in the standard \lcdm~ model \cite{kkv,moore} in contradiction
with observation around our galaxy.  Analytic calculations also 
suggest that galaxies should contain more satellites than observed
\cite{kwg}.  One may wonder that this discrepancy is simply due to 
the fact that these satellites are too dark to be observed.  The
situation, however, is not that simple because these satellites would
affect stability of galactic disks \cite{to}.  Hence the problem 
may not be solved
by an adjustment of the mass-to-light ratio.

Recently, Kamionkowski and Liddle \cite{kl} proposed that the  
dearth of substructure in galactic halos might be simply due to the lack of
small-scale power of initial fluctuations.  
They have shown that the discrepancy can be resolved if there is a
cutoff at the comoving wavenumber $k\simeq 4.5h {\rm Mpc}^{-1}\equiv k_c$
and if the amplitude of smaller-scale perturbation is more than, say, 
$10^{-2}$ times smaller than that on larger scale.
They have also argued that
such a spectrum could be produced if the slope of the inflaton's potential had a 
discontinuity.  While the spectrum of fluctuations produced in such a
broken-scale-invariance model has been elegantly studied by Starobinsky
\cite{staro}, there is no particle-physics motivation for considering
a scalar potential with its first derivative discontinuous.  It would be
much better to realize the same feature with a smooth potential.

In the present paper we propose an inflation model with the desired
spectrum in which the scalar potential is a smooth function of 
simple polynomials. 
We consider a kind of double inflation model \cite{double,chaonew2,ky}.
Scales probed by
large-scale structures and cosmic microwave background (CMB) radiation
leave the Hubble radius during the first inflation stage when almost
scale-invariant fluctuations with magnitude of order of $10^{-5}$ are
produced, while smaller scales leave the horizon during the second
inflation with the amplitude of perturbation smaller than $10^{-7}$.
Since the spectrum of fluctuations generated near the end of slow-roll 
inflation models such as new \cite{newinf} and chaotic \cite{chaoinf} 
inflation deviates from the scale-invariant one severely
\cite{chaonew2},
they cannot be a model for the first inflation.  Indeed the desired
cutoff scale is so close to those probed by large-scale structure and
CMB anisotropy that 
we must arrange the first inflation to terminate abruptly, in a time
scale smaller than the expansion time.  We adopt the hybrid
inflation model \cite{hybrid} for this purpose.  As shown in
\cite{hybrid}, hybrid inflation can end abruptly with a ``waterfall''
of a symmetry-breaking scalar field $\chi$, whose potential energy drives
inflation while its dynamics is controlled by another scalar field $\phi$.
We assume that the former field is also coupled to the inflaton field of
the second inflation and it sets an appropriate initial condition 
for the second inflation so that its number of $e$-folds is
controlled  in order to ensure the scale of the cutoff lies
at the right scale.

The reset of the paper is organized as follows.  In \S II we describe
our model Lagrangian and outline the scenario of cosmic evolution.  More
detailed analysis and constraints on the model parameters are given in 
\S III.  \S IV is devoted to discussion and conclusion.

\section{Model}

We consider a double hybrid inflation model.  We introduce two complex 
scalar fields, $\chione$ and $\chitwo$, which induce a series of
symmetry-breaking phase transitions, and two real scalar fields,
$\phione$ and $\phitwo$, that control the 
dynamics of $\chi$'s.  
The scalar part of our model Lagrangian is given as follows.
\beq
 \call \equiv \lmk \partial\chione\rmk^\dag \partial\chione
 +\half (\partial\phione)^2 + \lmk \partial\chitwo\rmk^\dag \partial\chitwo
 +\half (\partial\phitwo)^2 - V(\chione,\phione,\chitwo,\phitwo),
\eeq
\beq
  V(\chione,\phione,\chitwo,\phitwo)\equiv
\Vone (\chione,\phione)+
\Vtwo (\chitwo,\phitwo) + U(\chione,\phitwo), 
\eeq
\beqa
\Vone (\chione,\phione)&=& \half\mone^2\phione^2
+\frac{\lone}{2}\lmk |\chione|^2-\vone^2 \rmk^2
+\gone^2\phione^2|\chione|^2, \\
\Vtwo (\chitwo,\phitwo)&=& \half\mtwo^2\phitwo^2
+\frac{\ltwo}{2}\lmk |\chitwo|^2-\vtwo^2 \rmk^2
+\gtwo^2\phitwo^2|\chitwo|^2, \\
U(\chione,\phitwo)&=&-\half h^2\lmk  |\chione|^2-\vone^2 \rmk\phitwo^2
+h^2\lmk |\chione|^2-\vone^2 \rmk\sigma\phitwo, \label{U}
\eeqa
where $\gone,~\gtwo,~\lone,~\ltwo$ and $h$ are dimensionless coupling
constants, and $\mone,~\mtwo,~\vone,~\vtwo$ and $\sigma$ are constants with
mass dimension.  Obviously, $\Vone$ and $\Vtwo$ have the same structure 
with an appropriate form for hybrid inflation.  We assume that $\Vone$
is at higher energy scale than $\Vtwo$ and $U$.  
Then it drives the first inflation and
$\Vtwo$ and $U$ do not affect the dynamics of $\chione$ and $\phione$
in this epoch.

Under the above assumption, evolution of the universe proceeds as
follows.  The first and the second derivatives of $\Vone$ with respect to 
$\chione$ read
\beqa
  \frac{\partial \Vone}{\partial \chione^\dag}
 &=&\lone \lmk |\chione|^2-\vone^2 \rmk\chione +\gone^2\phione^2\chione,  \\ 
   \frac{\partial^2 \Vone}{\partial \chione \partial \chione^\dag}
 &\equiv& M^2_{\chione} =  
 \lone \lmk 2|\chione|^2-\vone^2 \rmk +\gone^2\phione^2,   
\eeqa
respectively.  Hence if $\gone^2\phione^2 > \lone\vone^2$ holds initially, $\chione$
has the global minimum at $\chione=0$.  We further take an 
initial condition such that $M_{\chione}$ is larger than the Hubble
parameter and $\chione$ settles to $\chione=0$ in the expansion time
scale.  Then the hybrid inflation is driven by the vacuum energy density
of $\chione$ with some contribution from the mass term of $\phione$, which
turns out to be small in the late stage of inflation and we neglect it
hereafter.  Thus the Hubble parameter, $H$, reads
\beq
   H^2 \cong \frac{8\pi}{3\mpl^2}V_1\cong
\frac{4\pi\lone\vone^4}{3\mpl^2}\equiv H_1^2.
\eeq
In this regime, the first and the second derivatives of $V$ with respect
to $\phitwo$ are given by,
\beqa
   \frac{\partial V}{\partial \phitwo}&=&\mtwo^2\phitwo
  +2\gtwo|\chitwo|^2\phitwo + h^2\vone^2(\phitwo - \sigma), \\ 
   \frac{\partial^2 V}{\partial \phitwo^2}
 &\equiv& M^2_{\phitwo} =\mtwo^2 + 2\gtwo^2|\chitwo|^2 +h^2\vone^2.
\eeqa  
Hence it has a minimum at
\beq
  \phitwo=\phitwom\equiv \frac{h^2(\vone^2-|\chione|^2)}{h^2(\vone^2-|\chione|^2)
+2\gtwo^2|\chitwo|^2+\mtwo^2}\sigma \cong \sigma
\eeq
to which $\phitwo$ is anchored provided $M^2_{\phitwo} > H_{1}^2$ at $\phitwo=\phitwom$.

Then, assuming  $\gtwo^2\sigma^2 > \ltwo\vtwo^2$, $\chitwo$ relaxes to
$\chitwo=0$ if $M_{\chitwo}^2 > H_1^2$.  Thus in this period we find $\chione=\chitwo=0$
and $\phitwo=\phitwom\cong\sigma$.
On the other hand, $\phione$ evolves slowly as
\beq
  \phione \propto \exp\lmk -\frac{\mone^2}{3H_1^2}H_1t\rmk,
\eeq
until it reaches the critical value $\phionec\equiv\sqrt{\lone}\vone/\gone$ when
$\chione=0$ becomes unstable to 
trigger the phase transition of $\chione$.  

The amplitude of curvature fluctuation generated on comoving scales that
leave the horizon during the first 
inflation is given by
\beq
  \Phi = \frac{3}{5}\frac{H_1}{2\pi\phione}\frac{3H_1^2}{\mone^2}, \label{yuragi1}
\eeq
in the matter dominated era, and the spectral index reads
\beq
  n_s=1+\frac{2\mone^2}{3H_1^2}.  \label{index} 
\eeq
The COBE observation \cite{cobe}, which probes the comoving scale about $e^{5.6}$
times larger than the desired cutoff scale, $2\pi/k_c$, constrains
$\Phi$ and $n_s$ as
\beqa
  \Phi &=& \frac{9H_1^3}{10\pi\phi_{1\ast}\mone^2} \simeq 3\times
10^{-5}, \label{SW} \\
  n_s &=& 1.2 \pm 0.3 \label{indexcobe},
\eeqa
with
\beq
  \phi_{1\ast}\equiv \phionec \exp\lmk \frac{5.6\mone^2}{3H_1^2}\rmk
  =\frac{\sqrt{\lone}\vone}{\gone}\exp\lmk \frac{5.6\mone^2}{3H_1^2}\rmk.
 \label{phionecobe}
\eeq

For $\phione < \phionec$, the minimum of 
$\chione$ is located on a circle with
\beq
|\chione|^2 = |\chionem|^2 \equiv \vone^2 - \frac{\gone^2}{\lone}\phione^2
=\frac{1}{\lone}M^2_{\chione}(\chione=0).
\eeq
But $\chione$ does not fall to this minimum immediately, because the
dynamics 
of $\chione$ 
is governed by quantum fluctuations intrinsic to an exponentially expanding spacetime 
during $|M^2_{\chione}(\chione=0)|\lsim H_1^2$ \cite{VOS}.  
Since we need to terminate the first
inflation abruptly, we request the time elapsed during 
$|M^2_{\chione}(\chione=0)|\lsim H_1^2$ is much smaller than the expansion time 
$H_1^{-1}$.  Then $\chione$ practically traces the evolution of $\chionem$, where
$\phione$ acquires a large mass, 
\beq
M_{\phione}^2=\mone^2 + \frac{\gone^2}{\lone}M^2_{\chione}(\chione=0).
\eeq
Thus, unless $\gone^2 \ll \lone$, $\phione$ rolls down to the origin 
immediately to terminate the 
phase transition and $\chione$ relaxes to $\chione = \vone$.

In this short
period, $\phitwom$ also shifts from $\sigma$ to zero.  However, it does not
change practically until $h^2(\vone^2 - |\chionem|^2)$ has become smaller than
$\mtwo^2$.  At this time $M_{\phitwo}^2$ has already decreased to be much smaller than
$H_1^2$, so that $\phitwo$ does not trace the evolution of $\phitwom$ but 
remains at $\sigma$ practically.

Thus after the first inflation both $\phione$ and $\chione$ dissipate
their energy through
oscillation while $\chitwo$ and $\phione$ are frozen at $0$ and $\sigma$, respectively.
Then the universe is soon dominated by the vacuum energy density of $\chitwo~(=0)$
and the second hybrid inflation sets in.  In this regime the amplitude of curvature
perturbation must be much smaller than $10^{-5}$ and the number of $e$-folds should be
around $N_2\simeq 50-55$, so that the cutoff of fluctuation spectrum lies at around
$k=4.5h {\rm Mpc}^{-1}$.  Here the precise value of $N_2$ depends on both the
scale of the second inflation and on the history of 
the universe after that, in particular, on whether a mini-inflation like thermal 
inflation occurs \cite{thermal}.  For definiteness, we take $N_2=50$ hereafter.

The second inflation ends as $\phitwo$ gets smaller than 
$\phitwoc\equiv \sqrt{\ltwo}\vtwo/\gtwo$ and $\chitwo$ becomes unstable.
Hence the maximum amplitude of curvature fluctuations generated in this stage is
given by
\beq
  \Phi = \frac{\gtwo H_2^3}{\pi\sqrt{\ltwo}\vtwo\mtwo^2},  \label{fluc2}
\eeq
measured at the horizon crossing in the radiation era after inflation.
It should be smaller than, say, $10^{-7}$ for our purpose.

The universe is reheated by the decay of $\chitwo$.  Since we do not
need to have tiny dimensionless coupling constants in hybrid inflation,
the reheat temperature can be much higher than the weak scale to
accommodate baryogenesis \cite{Dolgov}.

After the two inflations and the two phase transitions, we have two
networks of cosmic (global) strings, one on comoving horizon scale at
the end of inflation with energy scale $\vtwo$ and the other on
$e^{N_2}$ times larger scale with energy scale $\vone$.  These strings
are cosmologically harmless as long as $\vone$ and $\vtwo$ are smaller than
$7\times10^{14}$GeV \cite{global}.
This is why we have assumed $\chione$ and $\chitwo$ are complex fields
rather than real fields.  We would have encountered a fatal domain wall
problem in the latter case.

\section{Determination of model parameters}
Having outlined the evolutionary scenario of the universe in our model,
we now consider constraints on the model parameters.  
First  we study  constraints obtained from the first and  the second
inflation in order to produce the desired spectrum in turn.  Then we
consider the consistency conditions so that evolution of the universe
proceeds as outlined in the previous section and that only one field 
contributes to density fluctuations at each epoch.

\subsection{Constraints on $V_1$ from the first inflation }\label{first}

Define a dimensionless parameter
\beq
  \alphaone \equiv \frac{100\mone^2}{3H_1^2},  \label{alpha1}
\eeq
then we find
\beq
  \alphaone < 10,
\eeq
from  (\ref{index}) and the constraint on the spectral index
(\ref{indexcobe}). 

Then from the amplitude of fluctuations (\ref{SW}), the scale of the
first inflation is determined as
\beq
  \vone = 4.9\times 10^{-7}\gone^{-1}\alphaone e^{0.056\alphaone}\mpl,
  \label{v1}
\eeq
From $\vone \lsim 7\times 10^{14}$GeV, we find
\beq
  \gone > 8.4\times 10^{-3}\alphaone e^{0.056\alphaone}.
\eeq
(\ref{alpha1}) and (\ref{v1}) determines $\mone$ as
\beq
  \mone = 2.7\times 10^{-13}\lone^{\frac{1}{2}}\alphaone^{\frac{5}{2}}
  e^{0.11\alphaone}\mpl. \label{m1}
\eeq

From the dominance of the vacuum energy density of $\chione$, or
\beq
   \frac{\lone}{4}\vone^4 \gg \frac{1}{2}\mone^2\phi_{1\ast}^2,
\eeq
we find
\beq
  \gone \gg 7.5\times 10^{-4}\lone^{\frac{1}{4}}\alpha^{\frac{3}{4}}
  e^{0.056\alphaone}. 
\eeq

Since we are assuming that $\chione$ has been anchored at the origin by
the time $\phione$ reaches $\phi_{1\ast}$, we require
\beq
   M_{\chione}^2=\gone^2\phi_{1\ast}^2-\lone\vone^2+
  \frac{1}{2}h^2\sigma^2  > H_1^2
\eeq
which yields 
\beq
  \gone > 1.0\times 10^{-6}\alphaone e^{0.056\alphaone}
  \lmk e^{0.056\alphaone} -1\rmk^{-\frac{1}{2}},
\eeq
where contribution of $h^2\sigma^2/2$ is neglected.
As $\phione$ decrease to $\phionec$, $\chione=0$ becomes unstable and
a phase transition is triggered.  We require this transition should
occur abruptly, that is, the time elapsed from $M_\chione^2=H_1^2$ to
$M_\chione^2=-H_1^2$, which we denote by $\Delta t$, 
should be much smaller than the expansion time.  That is, 
\beq
  H_1\Delta t \cong \frac{3H_1^2}{2\mone^2}
\ln\left|\frac{\lone\vone^2+H_1^2}{\lone\vone^2-H_1^2}\right| 
\cong \frac{3H_1^4}{\lone\vone^2\mone^2}
  = \frac{4\pi}{3}\frac{\lone\vone^6}{\mone^2\mpl^4} \ll 1.
 \label{dt} 
\eeq
Here the second approximate equality applies because $\vone \lsim
7\times 10^{14}$GeV holds.
From (\ref{v1}) and (\ref{m1}), this yields
\beq
  H_1\Delta t = 1.0\times 10^{-10}\gone^{-2}\alphaone e^{0.11\alphaone}
\ll 1,
\eeq
or
\beq
  \gone \gg 1.0\times 10^{-5}\alphaone^{\frac{1}{2}}e^{0.056\alphaone}. 
\eeq

\subsection{Constraints on $V_2$ and $U$ from the second inflation}\label{second}
Since $\phitwo$ is proportional to $\exp\lmk{-\frac{\mtwo^2}{3H_2^2}H_2t}\rmk$
during the second  inflation, we find
\beq
  N_2=\frac{3H_2^2}{\mtwo^2}\ln\lmk\frac{\gtwo\sigma}{\sqrt{\ltwo}\vtwo}\rmk
 =\frac{2\pi\ltwo\vtwo^4}{\mtwo^2\mpl^2}
 \ln\lmk\frac{\gtwo\sigma}{\sqrt{\ltwo}\vtwo}\rmk.  \label{n2} 
\eeq
Let us take the logarithmic factor 
\beq
  \ln\lmk\frac{\gtwo\sigma}{\sqrt{\ltwo}\vtwo}\rmk=2, \label{log}
\eeq
for definiteness.  Then (\ref{fluc2}) reads
\beq
  \Phi=25\sqrt{\frac{2}{27\pi}}\frac{\gtwo\vtwo}{\mpl}
  \equiv 25\sqrt{\frac{2}{27\pi}}\times 10^{-8}\alphatwo
  =3.8\times 10^{-8}\alphatwo,
\eeq
with $\alphatwo$ being a constant of order of unity or smaller.
Then $\vtwo$ is determined as
\beq
  \vtwo = 10^{-8}\alphatwo\gtwo^{-1}\mpl.  \label{v2}
\eeq
Now $\vtwo \lsim 7\times 10^{14}$GeV implies 
\beq
\gtwo > 1.7\times 10^{-4}\alphatwo.
\eeq
We also find
\beq
  \mtwo = 5.0\times 10^{-17}\alphatwo^2\gtwo^{-2}\sqrt{\ltwo}\mpl,
\eeq
from (\ref{n2}) and 
\beq
  \sigma = 1.4\times 10^{-9}\alphatwo\gtwo^{-2}\sqrt{\ltwo}\mpl, \label{sigma}
\eeq
from (\ref{log}).
From the dominance of the vacuum energy of $\chitwo$, 
$\ltwo\vtwo^4/4 \gg \mtwo^2\phitwo^2/2$, we find
\beq
  \gtwo \gg 3.1\times 10^{-5}\alphatwo^{\frac{1}{2}},
\eeq
using (\ref{v2})-(\ref{sigma}).
Finally, the mass-squared of $\chitwo$ at the origin reads
\beq
  M^2_{\chitwo}(\chitwo=0,\phitwo=\sigma)=(e^4-1)\ltwo\vtwo^2
  =5.4\times 10^{-15}\ltwo\alphatwo^2\gtwo^{-2}\mpl^2,
\eeq
under our choice of the logarithmic factor (\ref{log}).  This should
be larger than the Hubble parameter-squared,
\beq
  H_2^2=\frac{2\pi\ltwo\vtwo^4}{2\mpl^2}=2.1\times
10^{-32}\ltwo\alphatwo^4 \gtwo^{-4}\mpl^2,
\eeq
so that $\chitwo$ is anchored at the origin.  This results in a very
mild constraint on $\gtwo$;
\beq
  \gtwo > 2.0\times 10^{-9}\alphatwo.
\eeq

\subsection{Consistency conditions}\label{consistency}

Here we study constraints on $V_2$ and $U$ from the first inflation and
those on $V_1$ from the second inflation, so that our scenario works
consistently. 

During the first inflation each term of $V_2$ and $U$ should be much
smaller than $\lone\vone^4/4$.  From $\lone\vone^4/4 \gg
\ltwo\vtwo^4/4$, we find
\beq
  \Gamma\equiv \frac{\lone\vone^4}{\ltwo\vtwo^4}=5.9\times 10^6
  \frac{\lone\gone^{-4}\alphaone^4 e^{0.22\alphaone}}
{\ltwo\gtwo^{-4}\alphatwo^4} \gg 1, \label{gamma}
\eeq
and from $\lone\vone^4/4 \gg h^2\vone^2\sigma^2/2$, we obtain
\beq
  \lone \gone^{-2}\alphaone^2 e^{0.11\alphaone} \gg 
  8.2\times 10^{-6}\ltwo\gtwo^{-4}\alphatwo^2h^2.    \label{42}
\eeq
We also require $M_\chitwo^2 > H_1^2$ and $M_\phitwo^2 > H_1^2$ at their
respective minima so that these fields are fixed there.  The former
constraint reads
\beq
  \ltwo\gtwo^{-2}\alphatwo^2 > 4.5\times 10^{-11}\lone \gone^{-4}
  \alphaone^4 e^{0.22\alphaone},
\eeq
while the latter implies
\beq
  h^2 > 1.0\times 10^{-12}\lone \gone^{-2}\alphaone^2 e^{0.11\alphaone}.
\eeq

On the other hand, $\chione$ and $\phione$ must be fixed at their ground
states during the second inflation.  Hence we require 
$M_\chione^2(\chione=\vone,\phione=0)>H_2^2$ and 
$M_\phione^2(\chione=\vone,\phione=0)>H_2^2$.  The former is
automatically satisfied because $M_\chione^2=\lone\vone^2 \gg H_1^2 \gg
H_2^2$, and the latter turns out to be a very weak constraint on $\ltwo$
and $\gtwo$;
\beq
  \ltwo\gtwo^{-4} < 2.3\times 10^{19}\alphaone^2\alphatwo^{-4}.  \label{45}
\eeq

\subsection{Determination of model parameters: An example}
 
Having shown all the necessary conditions we now demonstrate there is a
wide range of allowed region in the parameter space.
So far we have found four lower bounds on $\gone$ in \ref{first} and
three lower bounds on $\gtwo$ in \ref{second}.  On the other hand,
constraints obtained in \ref{consistency} are more complicated, so let
us take $\lone\gone^{-2}=1$ and $\ltwo\gtwo^{-2}=1$ for illustration.
These choices are also consistent with our assumption that $\gone^2$
$(\gtwo^2)$  is not much smaller than $\lone$ $(\ltwo)$.

Then (\ref{42}) through (\ref{45}) read, respectively,
\beqa
  \gtwo &\gg & 2.9\times
10^{-3}\alphaone^{-1}e^{-0.056\alphaone}\alphatwo h, \\
  \gone &\gg & 6.7\times 10^{-6}\alphaone^2 e^{0.11\alphaone}
\alphatwo^{-1}, \\
  h &>& 1.0\times 10^{-6}\alphaone e^{0.056\alphaone}, \\
\gtwo &>& 2.1\times 10^{-10}\alphaone^{-1}\alphatwo^2.
\eeqa
   
All of the constraints shown so far are so mild that it is easy to choose 
a consistent set of parameter values.  Here we merely give an example.
Let us take $\Gamma$ not so big, say $\Gamma=25$, so that the second inflation
sets in soon after the end of the first one and that the intermediate period
is unimportant.  Let us further take $\alphaone=0.1$ and $\alphatwo=1$.
Then $\Gamma=25$ implies $\gtwo=0.21\gone$ under the choice 
$\lone\gone^{-2}=\ltwo\gtwo^{-2}=1$.

Now all the constraints turn out to be those on $\gone$ and $h$, which are 
summarized as
\beqa 
   \gone &>& 8.4\times 10^{-4}, \\
   \gone &\gg& 1.5\times 10^{-4}, \\
   \gone &\gg& 0.14h, \\
   h     &\gg& 1.0\times 10^{-7}.
\eeqa
The scale of inflation then read
\beqa
   \vone &=& 5.9\times 10^{11}\gone^{-1} {\rm GeV}, \\
   \vtwo &=& 5.7\times 10^{11}\gone^{-1} {\rm GeV},
\eeqa
together with $\ltwo =4.4\times 10^{-2}\lone = 
4.4\times 10^{-2}\gone^2$.  Other mass scales of the model are given by
\beqa
  \mone &=& 10\gone {\rm TeV},\\
  \mtwo &=& 2.9\gone {\rm TeV},\\
  \sigma &=& 8.0\times 10^{10}\gone^{-1} {\rm GeV}.
\eeqa 

If we take $\gone=0.1$, for example, all the dimensionless coupling
constants take natural values and $\vone$ and $\vtwo$ fall on the typical
symmetry breaking scale for the axion \cite{axion}.

\section{Conclusion}

In the present paper we have presented a simple model of double
inflation that can realize a spectrum of density fluctuation with a
small-scale cutoff at the comoving wavenumber at $k_c\simeq 4.5h {\rm
Mpc}^{-1}$ in order to explain the dearth of substructure in galactic
halos.  Since the spectrum of density fluctuations generated in the late
stage of new and chaotic inflation models deviates from the
scale-invariant one significantly, we adopted the hybrid inflation model
for the first inflation which is responsible for fluctuations on scales
probed by large-scale structures and CMB anisotropy.  We have also made
use of the hybrid inflation model for the second inflation because it is
easy to control the energy scale of inflation in this model.  

Since each hybrid inflation requires at least two scalar fields, our model
includes four scalar fields as a double hybrid inflation model.  In this
sense our model is somewhat complicated, but our scalar potential
consists of a polynomial up to only fourth order and all the parameters
take natural magnitudes.  
There are, however, two unwanted features in this model.  In concluding
the paper, we point them
out and mention possible resolutions in turn.

One is the problem of initial condition.  It has been discussed that
only a tiny fraction in the field configuration space can give rise to
successful hybrid inflation, because $\phi$  rolls
down the potential rapidly if $\chi$ is too large initially \cite{tetradis}.
The resolution to this problem has recently been proposed \cite{ML}, 
which introduces more fields or brane cosmology.

The other is the linear interaction term of $\phitwo$ in (\ref{U}),
which is required in order to shift the potential minimum of $\phitwo$
out of $\phitwo=0$ during the first inflation.  There is no
particle-physics motivation to introduce this type of direct linear
coupling.  
But such a term is effectively induced naturally in more complicated classes of
supergravity inflation models \cite{ky}.

The purpose of the present paper was to make  the simplest
possible toy model to realize the desired spectral shape of primordial
density fluctuations in order to stress the essential ingredients of the model.  
In a forthcoming paper we present a more
realistic supergravity model to realize the same spectral feature.

\acknowledgements{The author is grateful to F.\ Takahara for useful
communications.  This work was partially supported by the Monbusho
Grant-in-Aid, Priority Area ``Supersymmetry and Unified Theory of
Elementary Particles''(\#707) and the Monbusho Grant-in-Aid for
Scientific Research No.\ 11740146.}

\end{document}